\documentclass[11pt]{cernrep}
\usepackage{graphicx}
\usepackage{cite,./mcite}

\newcommand{\herwig}{\textsc{herwig}}
\newcommand{\hpp}{\textrm{Herwig++}}
\newcommand{\ThePEG}{\textrm{ThePEG}}
\newcommand{\jimmy}{\textsc{jimmy}}

\begin{document}
\title{Herwig++ Status Report\footnote{{}\hspace{0.5em}to appear in the proceedings of
    the HERA and the LHC workshop.}
\hfill\parbox[b]{3cm}{\rm \footnotesize\raggedleft 
CP3-08-42\\DCTP/08/134\\IPPP/08/67\\KA-TP-20-2008\\MCnet/08/10}}

\author{M.~B\"ahr$^{\rm a}$, S.~Gieseke$^{\rm a}$, M.A.~Gigg$^{\rm b}$, D.~Grellscheid$^{\rm b}$, K.~Hamilton$^{\rm c}$,
O.~Latunde-Dada$^{\rm d}$, S.~Pl\"atzer$^{\rm a}$, P.~Richardson$^{\rm b,e}$, M.H.~Seymour$^{\rm e,f}$,
A. Sherstnev$^{\rm d}$, J.~Tully$^{\rm b}$, B.R.~Webber$^{\rm d}$}
\institute{
  $^{\rm a}$\,Institut f\"ur Theoretische Physik, Universit\"at Karlsruhe,\\
  $^{\rm b}$\,IPPP, Department of Physics, Durham University,\\
  $^{\rm c}$\,Centre for Particle Physics and Phenomenology, 
  Universit\'e Catholique de Louvain,\\
  $^{\rm d}$\,Cavendish Laboratory, University of Cambridge,\\
  $^{\rm e}$\,Physics Department, CERN,\\
  $^{\rm f}$\,School of Physics and Astronomy, University of Manchester.
  \footnote{{}\hspace{0.5em}\texttt{herwig@projects.hepforge.org}}
}

\maketitle

\begin{abstract}
  \hpp{} is the successor of the event generator \herwig{}.  In its
  present version 2.2.1 it provides a program for full LHC event
  generation which is superior to the previous program in many respects.
  We briefly summarize its features and describe present work and some
  future plans.\\  
%\textbf{TODO: bib style should give arXiv numbers.} 

\end{abstract}

\section{Introduction}
\label{sec:intro}
With the advent of the LHC era it was decided to completely rewrite the
general purpose event generator \herwig{}
\cite{Corcella:2000bw,Corcella:2002jc} in C++ under the name \hpp{},
based on the package \ThePEG \cite{Lonnblad:2006pt,Bertini:2000uh}.  The
goal is not only to provide a simple replacement of \herwig{} but to
incorporate physics improvements as well \cite{Gieseke:2002sg}.  From
2001 until now \hpp{} has been continuously developed and
extended\cite{Gieseke:2003hm,Gieseke:2006rr,Gieseke:2006ga,Bahr:2007ni,Bahr:2008tx}.
The current version is 2.2.1, cf.\ \cite{Hppurl}.  The physics
simulation of the current version is more sophisticated than the one of
Fortran \herwig{} in many respects.  In this report we will briefly
summarize the status of the different aspects of the simulation.  These
are the hard matrix elements available, initial and final state parton
showers, the hadronization, hadronic decays and the underlying event.
We conclude with an outlook to planned future improvements.

\section{Physics simulation steps}
\label{sec:phys}

\subsection{Matrix elements}
\label{sec:me}

The event generation begins with the hard scattering of incoming
particles or partons in the case of hadronic collisions.  We have
included a relatively small number of hard matrix elements.  These
include $e^+e^-$ annihilation to $q\bar q$ pairs or simply to $Z^0$
bosons and deep inelastic scattering.  In addition there is the
Higgsstrahlung process $e^+e^-\to h^0Z^0$. For hadron--hadron collisions we
have the QCD $2\to 2$ processes including heavy quark production. 
For colourless final states we have the following matrix elements, 
\begin{displaymath}
  hh\to (\gamma, Z^0)\to \ell^+\ell^-\ , \quad
  hh\to W^\pm \to \ell^\pm\nu_\ell(\bar \nu_\ell)\ , \quad
  hh\to h^0\ , \quad
  hh\to h^0Z^0\ , \quad  
  hh \to \gamma\gamma\ .
\end{displaymath}
We also provide matrix elements for processes with additional jets in
the final state, like
\begin{displaymath}
  hh\to (\gamma, Z^0, W^\pm) +{\rm jet}\ , \quad
  hh\to h^0 + {\rm jet}\ .
\end{displaymath}
In addition, there are matrix elements for perturbative decays of the
top quark, which will be simulated including spin correlations (see
below).  There will be some more matrix elements added in future
versions, e.g.\ for $hh\to qqh^0$.  Despite the rather small number of
matrix elements, there is no real limitation to the processes that may
be simulated with \hpp{}.  In practice, one may use any matrix element
generator to generate a standard event file \cite{Alwall:2006yp} which
in turn can be read and processed by \hpp{}.

For processes with many legs in the final state we follow a different
strategy.  When the number of legs becomes large --- typically larger
than 6--8 particles in the final state --- it will be increasingly
difficult to achieve an efficient event generation of the full matrix
element.  For these situations we have a generic framework to build up
matrix elements for production and decays of particles in order to
approximate any tree level matrix element as a simple production process
with subsequent two or three body decays.  This is a good approximation
whenever the widths of the intermediate particles are small.  The spin
correlations among these particles can be restored with the algorithm
described in \cite{Richardson:2001df}.  Also finite width effects are
taken into account \cite{Gigg:2008yc}.  The full simulation of several
processes of many models for physics beyond the standard model (MSSM,
UED, Randall--Sunrum model) is thus possible in \hpp{}
\cite{Gigg:2007cr}.  Here, all necessary matrix elements for production
and decay processes are constructed automatically from a model file.

\subsection{Parton Showers and matching with matrix elements}
\label{sec:shower}

After the hard process has been generated, typically at a large scale
$\sim 100\,$GeV--1\,TeV, the coloured particles in the process radiate a
large number of additional partons, predominantly gluons.  As long as
these are resolved by a hard scale of $\sim 1\,$GeV this is simulated
with a coherent branching algorithm, as outlined in
\cite{Gieseke:2003rz} which generalizes the original algorithm
\cite{Marchesini:1983bm,Webber:1983if,Marchesini:1987cf} used in
\herwig{}.  The main improvements with respect to the old algorithm are
boost invariance along the jet axis, due to a covariant formulation, and
the improved treatment of radiation off heavy quarks.  We are using
mass--dependent splitting functions and a description of the kinematics
that allows us to dynamically generate the dead--cone effect.  In
addition to initial and final state parton showers there are also parton
showers in the decay of heavy particles, the top quark in our case.

When extrapolating to hard, wide--angle emissions, the parton shower
description is not sufficiently accurate in situations where observables
depend on large transverse momenta in the process.  In these cases we
supply so--called hard matrix element corrections that describe the
hardest parton emission, usually a hard gluon, with the full matrix
element for the process that includes that extra parton.  In order to
consistently describe the whole phase space one has to apply soft matrix
element corrections.  Matrix element corrections are available for
Drell--Yan type processes, Higgs production in $gg$ fusion and $e^+e^-$
annihilation to $q\bar q$--pairs.  In addition, we apply a matrix
element correction in top--quark decays \cite{Hamilton:2006ms}.

From the point of view of perturbation theory, the hard matrix element
correction is only one part of the next--to--leading order (NLO)
correction to the Born matrix element.  The full NLO calculation also
includes the virtual part with the same final state as the Born
approximation.  When trying to match NLO calculations and parton shower
algorithms systematically, we have to avoid double counting of the real
emission contributions.  Two systematic approaches are being
successfully discussed and applied in event generators: MC@NLO
\cite{Frixione:2002ik,Frixione:2003ei,Frixione:2006gn} and the POWHEG
approach \cite{Nason:2004rx,Frixione:2007vw}.  In \hpp{} we have
included working examples of matching in both approaches. The MC@NLO
method, adopted to \hpp{} is described in \cite{LatundeDada:2007jg}.
Whereas the POWHEG method has already been applied for several processes
in $e^+e^-$ annihilation \cite{LatundeDada:2006gx,LatundeDada:2008bv}
and also for Drell--Yan production \cite{Hamilton:2008pd}.  Parts of
these implementations will become available in future releases.

Another viable possibility to improve the description of QCD radiation
in the event generation is the matching to multiple tree--level matrix
elements, that describe the radiation of $n$ additional jets with
respect to the Born level.  Theoretically most consistent is the CKKW
approach \cite{Catani:2001cc} which has been studied in the context of
an angular ordered parton shower in \cite{PlaetzerThesis}.

\subsection{Hadronization and decays}
\label{sec:had}

The hadronization model in \hpp{} is the cluster hadronization model
which has not been changed much from its predecessor in \herwig{}.
After the parton shower, all gluons are split nonperturbatively into
$q\bar q$ pairs.  Then, following the colour history of the parton
cascade, all colour triplet--antitriplet pairs are paired up in
colourless clusters which still carry all flavour and momentum
information of the original partons.  While these are heavier than some
threshold mass they will fission into lighter clusters until all
clusters are sufficiently light.  These light clusters will then decay
into pairs of hadrons.

The hadrons thus obtained are often heavy resonances that will
eventually decay on timescales that are still irrelevant for the
experiment.  These hadronic decays have been largely rewritten and are
modeled in much greater detail in \hpp{}.  While in \herwig{} they were
often simply decayed according to the available phase space only, we now
take into account more experimental information, like form factors, that
allow for a realistic modeling of decay matrix elements
\cite{Grellscheid:2007tt,Bahr:2008pv}.  In a major effort, a large
fraction of the decay channels described in the particle data book
\cite{Amsler:2008zz} have been included into \hpp{}.

\subsection{Underlying event}
\label{sec:ue}

The underlying event model of \hpp{} is a model for multiple hard
partonic interactions, based on an eikonal model, similar to \jimmy{}
\cite{Butterworth:1996zw}.  In addition to the signal process there are
a number of additional QCD scatters, including full parton showers, that
contribute to the overall hadronic activity in the final state and
eventually also give rise to a (relatively soft) jet substructure in the
underlying event.  The model has two important parameters, one parameter
$\mu$, describing the spatial density of partonic matter in the
colliding protons. Secondly, there is one cut off parameter $p_{\perp,
  {\rm min}}$ that gives a lower bound on the differential cross section
for QCD $2\to 2$ jet production.  The model has been carefully tuned to
Tevatron data \cite{Bahr:2008dy}. Further possible bounds on the model
parameters have been studied in \cite{Bahr:2008wk}.  An alternative
modeling of the underlying event on the basis of the UA5 model
\cite{Alner:1986is} is also available for historic reasons.

Currently, the multiple partonic interaction model is limited to hard
scattering while a soft component is simply not present.  For a
realistic simulation of minimum bias events a soft component is,
however, very important.  An extension into the soft region, allowing us
the simulation of minimum bias events is currently being studied and is
likely to be included in the next release of \hpp{}.

\section{Availability}
\label{sec:avail}

The latest version of \hpp{} is always available from hepforge:
\begin{quote}
  \texttt{http://projects.hepforge.org/herwig}
\end{quote}
There one can also find wiki pages to help with questions concerning
installation, changing particular parameters and other frequently asked
questions.  The installation process is straightforward on any modern
variant of linux.  The physics details of the program are now documented
in great detail in our manual \cite{Bahr:2008pv}.  The pdf version of
the manual contains addional links to the online documentation of the
code.  All important parameters have been carefully tuned to a wealth of
available data and the code is shipped with default paramters that give
the best overall description of the data that we have tuned to. Details
of the tune can also be found in the manual \cite{Bahr:2008pv}.

\section*{Acknowledgments}

This work has been supported in part by the EU Marie Curie Research and
Training Network MCnet under contract MRTN-CT-2006-035606. 

%------------------------------------------------------------------------------
%       Bibliography
%------------------------------------------------------------------------------
\bibliographystyle{heralhc} 
{\raggedright
\bibliography{heralhc}

\providecommand{\href}[2]{#2}\begingroup\raggedright\begin{thebibliography}{10}

\bibitem{Corcella:2000bw}
G.~Corcella {\em et.~al.}, {\it {HERWIG 6: An event generator for hadron
  emission reactions with interfering gluons (including supersymmetric
  processes)}},  {\em JHEP} {\bf 01} (2001) 010,
  [\href{http://xxx.lanl.gov/abs/hep-ph/0011363}{{\tt hep-ph/0011363}}].

\bibitem{Corcella:2002jc}
G.~Corcella {\em et.~al.}, {\it {HERWIG 6.5 release note}},
  \href{http://xxx.lanl.gov/abs/hep-ph/0210213}{{\tt hep-ph/0210213}}.

\bibitem{Lonnblad:2006pt}
L.~L\mbox{\"{o}}nnblad, {\it {ThePEG, Pythia7, Herwig++ and Ariadne}},  {\em
  Nucl. Instrum. Meth.} {\bf A559} (2006) 246--248.

\bibitem{Bertini:2000uh}
M.~Bertini, L.~L\mbox{\"{o}}nnblad, and T.~Sj\mbox{\"{o}}strand, {\it {Pythia
  version 7-0.0: A proof-of-concept version}},  {\em Comput. Phys. Commun.}
  {\bf 134} (2001) 365--391,
  [\href{http://xxx.lanl.gov/abs/hep-ph/0006152}{{\tt hep-ph/0006152}}].

\bibitem{Gieseke:2002sg}
S.~Gieseke, {\it {Event generators: New developments}},
  \href{http://xxx.lanl.gov/abs/hep-ph/0210294}{{\tt hep-ph/0210294}}.

\bibitem{Gieseke:2003hm}
S.~Gieseke, A.~Ribon, M.~H. Seymour, P.~Stephens, and B.~Webber, {\it {Herwig++
  1.0: An event generator for e+ e- annihilation}},  {\em JHEP} {\bf 02} (2004)
  005, [\href{http://xxx.lanl.gov/abs/hep-ph/0311208}{{\tt hep-ph/0311208}}].

\bibitem{Gieseke:2006rr}
S.~Gieseke {\em et.~al.}, {\it {Herwig++ 2.0 beta release note}},
  \href{http://xxx.lanl.gov/abs/hep-ph/0602069}{{\tt hep-ph/0602069}}.

\bibitem{Gieseke:2006ga}
S.~Gieseke {\em et.~al.}, {\it {Herwig++ 2.0 release note}},
  \href{http://xxx.lanl.gov/abs/hep-ph/0609306}{{\tt hep-ph/0609306}}.

\bibitem{Bahr:2007ni}
M.~B\mbox{\"{a}}hr {\em et.~al.}, {\it {Herwig++ 2.1 Release Note}},
  \href{http://xxx.lanl.gov/abs/0711.3137}{{\tt arXiv:0711.3137}}.

\bibitem{Bahr:2008tx}
M.~B\mbox{\"{a}}hr {\em et.~al.}, {\it {Herwig++ 2.2 Release Note}},
  \href{http://xxx.lanl.gov/abs/0804.3053}{{\tt arXiv:0804.3053}}.

\bibitem{Hppurl}
\texttt{http://projects.hepforge.org/herwig}.

\bibitem{Alwall:2006yp}
J.~Alwall {\em et.~al.}, {\it {A standard format for Les Houches event files}},
   {\em Comput. Phys. Commun.} {\bf 176} (2007) 300--304,
  [\href{http://xxx.lanl.gov/abs/hep-ph/0609017}{{\tt hep-ph/0609017}}].

\bibitem{Richardson:2001df}
P.~Richardson, {\it {Spin correlations in Monte Carlo simulations}},  {\em
  JHEP} {\bf 11} (2001) 029,
  [\href{http://xxx.lanl.gov/abs/hep-ph/0110108}{{\tt hep-ph/0110108}}].

\bibitem{Gigg:2008yc}
M.~A. Gigg and P.~Richardson, {\it {Simulation of Finite Width Effects in
  Physics Beyond the Standard Model}},
  \href{http://xxx.lanl.gov/abs/0805.3037}{{\tt arXiv:0805.3037}}.

\bibitem{Gigg:2007cr}
M.~Gigg and P.~Richardson, {\it {Simulation of beyond standard model physics in
  Herwig++}},  {\em Eur. Phys. J.} {\bf C51} (2007) 989--1008,
  [\href{http://xxx.lanl.gov/abs/hep-ph/0703199}{{\tt hep-ph/0703199}}].

\bibitem{Gieseke:2003rz}
S.~Gieseke, P.~Stephens, and B.~Webber, {\it {New formalism for QCD parton
  showers}},  {\em JHEP} {\bf 12} (2003) 045,
  [\href{http://xxx.lanl.gov/abs/hep-ph/0310083}{{\tt hep-ph/0310083}}].

\bibitem{Marchesini:1983bm}
G.~Marchesini and B.~R. Webber, {\it {Simulation of QCD Jets Including Soft
  Gluon Interference}},  {\em Nucl. Phys.} {\bf B238} (1984) 1.

\bibitem{Webber:1983if}
B.~R. Webber, {\it {A QCD Model for Jet Fragmentation Including Soft Gluon
  Interference}},  {\em Nucl. Phys.} {\bf B238} (1984) 492.

\bibitem{Marchesini:1987cf}
G.~Marchesini and B.~R. Webber, {\it {Monte Carlo Simulation of General Hard
  Processes with Coherent QCD Radiation}},  {\em Nucl. Phys.} {\bf B310} (1988)
  461.

\bibitem{Hamilton:2006ms}
K.~Hamilton and P.~Richardson, {\it {A simulation of QCD radiation in top quark
  decays}},  {\em JHEP} {\bf 02} (2007) 069,
  [\href{http://xxx.lanl.gov/abs/hep-ph/0612236}{{\tt hep-ph/0612236}}].

\bibitem{Frixione:2002ik}
S.~Frixione and B.~R. Webber, {\it {Matching NLO QCD computations and parton
  shower simulations}},  {\em JHEP} {\bf 06} (2002) 029,
  [\href{http://xxx.lanl.gov/abs/hep-ph/0204244}{{\tt hep-ph/0204244}}].

\bibitem{Frixione:2003ei}
S.~Frixione, P.~Nason, and B.~R. Webber, {\it {Matching NLO QCD and parton
  showers in heavy flavour production}},  {\em JHEP} {\bf 08} (2003) 007,
  [\href{http://xxx.lanl.gov/abs/hep-ph/0305252}{{\tt hep-ph/0305252}}].

\bibitem{Frixione:2006gn}
S.~Frixione and B.~R. Webber, {\it {The MC@NLO 3.3 event generator}},
  \href{http://xxx.lanl.gov/abs/hep-ph/0612272}{{\tt hep-ph/0612272}}.

\bibitem{Nason:2004rx}
P.~Nason, {\it {A new method for combining NLO QCD with shower Monte Carlo
  algorithms}},  {\em JHEP} {\bf 11} (2004) 040,
  [\href{http://xxx.lanl.gov/abs/hep-ph/0409146}{{\tt hep-ph/0409146}}].

\bibitem{Frixione:2007vw}
S.~Frixione, P.~Nason, and C.~Oleari, {\it {Matching NLO QCD computations with
  Parton Shower simulations: the POWHEG method}},  {\em JHEP} {\bf 11} (2007)
  070, [\href{http://xxx.lanl.gov/abs/0709.2092}{{\tt arXiv:0709.2092}}].

\bibitem{LatundeDada:2007jg}
O.~Latunde-Dada, {\it {Herwig++ Monte Carlo At Next-To-Leading Order for e+e-
  annihilation and lepton pair production}},  {\em JHEP} {\bf 11} (2007) 040,
  [\href{http://xxx.lanl.gov/abs/0708.4390}{{\tt arXiv:0708.4390}}].

\bibitem{LatundeDada:2006gx}
O.~Latunde-Dada, S.~Gieseke, and B.~Webber, {\it {A positive-weight
  next-to-leading-order Monte Carlo for e+ e- annihilation to hadrons}},  {\em
  JHEP} {\bf 02} (2007) 051,
  [\href{http://xxx.lanl.gov/abs/hep-ph/0612281}{{\tt hep-ph/0612281}}].

\bibitem{LatundeDada:2008bv}
O.~Latunde-Dada, {\it {Applying the POWHEG method to top pair production and
  decays at the ILC}},  \href{http://xxx.lanl.gov/abs/0806.4560}{{\tt
  arXiv:0806.4560}}.

\bibitem{Hamilton:2008pd}
K.~Hamilton, P.~Richardson, and J.~Tully, {\it {A Positive-Weight
  Next-to-Leading Order Monte Carlo Simulation of Drell-Yan Vector Boson
  Production}},  \href{http://xxx.lanl.gov/abs/0806.0290}{{\tt
  arXiv:0806.0290}}.

\bibitem{Catani:2001cc}
S.~Catani, F.~Krauss, R.~Kuhn, and B.~R. Webber, {\it {QCD matrix elements +
  parton showers}},  {\em JHEP} {\bf 11} (2001) 063,
  [\href{http://xxx.lanl.gov/abs/hep-ph/0109231}{{\tt hep-ph/0109231}}].

\bibitem{PlaetzerThesis}
S.~Pl\mbox{\"{a}}tzer, {\it {Diploma Thesis }}, . Universit\mbox{\"{a}}t
  Karlsruhe, 2006.

\bibitem{Grellscheid:2007tt}
D.~Grellscheid and P.~Richardson, {\it {Simulation of Tau Decays in the
  Herwig++ Event Generator}},  \href{http://xxx.lanl.gov/abs/0710.1951}{{\tt
  arXiv:0710.1951}}.

\bibitem{Bahr:2008pv}
M.~B\mbox{\"{a}}hr {\em et.~al.}, {\it {Herwig++ Physics and Manual}},
  \href{http://xxx.lanl.gov/abs/0803.0883}{{\tt arXiv:0803.0883}}.

\bibitem{Amsler:2008zz}
{\bf Particle Data Group} Collaboration, C.~Amsler {\em et.~al.}, {\it {Review
  of particle physics}},  {\em Phys. Lett.} {\bf B667} (2008) 1.

\bibitem{Butterworth:1996zw}
J.~M. Butterworth, J.~R. Forshaw, and M.~H. Seymour, {\it {Multiparton
  interactions in photoproduction at HERA}},  {\em Z. Phys.} {\bf C72} (1996)
  637--646, [\href{http://xxx.lanl.gov/abs/hep-ph/9601371}{{\tt
  hep-ph/9601371}}].

\bibitem{Bahr:2008dy}
M.~B\mbox{\"{a}}hr, S.~Gieseke, and M.~H. Seymour, {\it {Simulation of multiple
  partonic interactions in Herwig++}},  {\em JHEP} {\bf 07} (2008) 076,
  [\href{http://xxx.lanl.gov/abs/0803.3633}{{\tt arXiv:0803.3633}}].

\bibitem{Bahr:2008wk}
M.~B\mbox{\"{a}}hr, J.~M. Butterworth, and M.~H. Seymour, {\it {The Underlying
  Event and the Total Cross Section from Tevatron to the LHC}},
  \href{http://xxx.lanl.gov/abs/0806.2949}{{\tt arXiv:0806.2949}}.

\bibitem{Alner:1986is}
{\bf UA5} Collaboration, G.~J. Alner {\em et.~al.}, {\it {The UA5 High-Energy
  anti-p p Simulation Program}},  {\em Nucl. Phys.} {\bf B291} (1987) 445.

\end{thebibliography}\endgroup
}
\end{document}